\documentclass[12pt,epsf]{article}
\usepackage{epsfig}

\renewcommand{\thanks}[1]{\footnote{#1}} % Use this for footnotes

\newcommand{\be}{\begin{equation}}
\newcommand{\ee}{\end{equation}}
\newcommand{\bea}{\begin{eqnarray}}
\newcommand{\eea}{\end{eqnarray}}

\begin{document}

\pagestyle{empty}

\bigskip\bigskip
\begin{center}
{\bf \large Thermal Evolution of a Dual Scale Cosmology}
\end{center}

\begin{center}
James Lindesay\footnote{e-mail address, jlslac@slac.stanford.edu} \\
Computational Physics Laboratory \\
Howard University,
Washington, D.C. 20059 
\end{center}

\begin{center}
{\bf Abstract}
\end{center}
Previous work developed a space-time metric with
two cosmological scales; one that conveniently
describes the classical evolution of the dynamics, and
the other describing a scale associated with
macroscopic quantum aspects like vacuum energy. 
The present work expands upon the dynamics of
these scales to demonstrate the usefulness of
these coordinates for describing early and late
time behaviors of our universe.  A convenient
parameter, the fraction of classical energy density,
is introduced as a means to parameterize the various
early time models for the microscopic input.

\setcounter{equation}{0}
\section{Introduction}
\indent

The appearance of multiple scales of cosmological
relevance from observational evidence such as supernovae
accelerations\cite{TypeIa} and CMB fluctuations\cite{WMAP}
motivates the development of models with scale dependence
in the early and late times, but spatial scale invariance during
intermediate times.  A metric with coordinates that conveniently
express such dependencies was developed in a previous
paper\cite{FRWdS}.  The present paper will further develop the
usefulness of this description of space-time for exploring the
dynamics of our cosmology.
 
It will be assumed that the dynamics can be 
accurately described using the Einstein equation without
a cosmological constant during the period
under consideration:
\be
G_{\mu \nu} \equiv \mathbf{R}_{\mu \nu} -
{1 \over 2} g_{\mu \nu} \mathbf{R} = - 
{8 \pi G_N \over c^4} T_{\mu \nu} ,
\label{EinsteinEqn}
\ee
where the energy-momentum tensor
takes the form of an ideal fluid
\be
T_{\mu \nu} = P \: g_{\mu \nu} + (\rho + P) u_\mu u_\nu.
\label{Tmunu}
\ee
This fluid form considerably constrains the possible forms
of the geometric dynamics of the cosmology.

\subsection{A hybrid metric}
\indent

Previous work\cite{FRWdS} has motivated the use of the following hybrid metric
well suited for the exploration of the early and late time behaviors of the cosmology:
\be
\begin{array}{r}
g_{\mu \nu} =
 -  c^2 dt^2 + R^2(ct) \left( dr - {r \over R_v (ct)} c dt  \right )^2  \\
+ R^2 (ct) \left ( r^2 d \theta ^2 + r^2 sin^2 \theta d\phi ^2   \right ) .
\end{array}
\label{FRWdSmetric}
\ee
This metric is convenient in its explicit inclusion of a dynamical parameters
that can generate a deSitter geometry in late times.
The parameter $R_v$ provides an intuitive scale that describes
the evolution of the cosmological (quantum) vacuum modes, while
the parameter $R$ describes the evolution of the (classical)
thermal state of the cosmology.  
The hydrodynamic parameters can be immediately calculated using
Eq. \ref{EinsteinEqn}:
\be
\begin{array}{l}
\rho= {3 c^4 \over 8 \pi G_N} \left (
{1 \over R_v}  + {\dot{R} \over R }    \right )^2 \equiv
{3 c^4 \over 8 \pi G_N} \left (
 { \dot{\mathcal{R}} \over \mathcal{R} } \right )^2, \\ \\
P  + \rho = - {c^4 \over 4 \pi G_N} {d \over dct} \left (
{1 \over R_v} + {\dot{R} \over R }  \right ) =
- {c^4 \over 4 \pi G_N} {d \over dct} \left (
{ \dot{\mathcal{R}} \over \mathcal{R} }  \right ) .
\end{array}
\label{FRWdSdensity}
\ee
The dynamics in Eq. \ref{FRWdSdensity} can be expressed
solely in terms of the energy content:
\be
{d \over d ct} \rho = -\sqrt{{24 \pi G_N \rho  \over c^4 }} (P + \rho) ,
\label{rhodynamics}
\ee
which, if the energy content is partitioned in terms of the
gravitational vacuum energy and thermal energy
$\rho=\rho_v + \rho_{thermal}$, and the equation of
state takes the usual form $P=w \rho$, gives an equation
for the evolution of the energy density of the form
\be
{d \over d ct} \rho = -  (1+w) \rho_{thermal}
\sqrt{{24 \pi G_N  \over c^4 } \rho } .
\label{thermalrhodynamics}
\ee

\subsection{Temporal and spatial scales}
\indent

The early time microscopic dynamics will define a temporal scale
associated with the thermalization of any quantum coherent
initial state of the cosmology.  This time scale is expected
to be associated with the small spatial scale cutoff of the macroscopic
gravitational physics, and will be referred to as $\tau_{UV}$.  Early
temporal evolution is defined by times less than or comparable to
this scale.  Similarly, late time behavior is characterized by the
domination of dark energy, and hence will be determined in
terms of the temporal scale $\tau_\Lambda$.  For the
cosmology of interest, $\tau_\Lambda >>> \tau_{UV}$.  The
intermediate, thermal evolution occurs for periods
between these extremes, when the energy content is dominated
by radiation and/or matter.

Generally,
the final state scale $R_v$ need not be a fixed constant.  However,
for the dual scale cosmology discussed here the scale parameters
will be assumed to behave as follows:
\be
\begin{array}{c l}
R_I \Leftarrow R_v  (ct) \Rightarrow R_\Lambda  & 
for \quad 0 \leftarrow t / \tau_{UV} \rightarrow \infty \\
R_I \Leftarrow R(ct) \Rightarrow R_\Lambda  & 
for \quad 0 \leftarrow t / \tau_\Lambda \rightarrow \infty \\
\rho_I \Leftarrow \rho (ct) = \rho_v + \rho_{thermal} \Rightarrow \rho_\Lambda  & 
for \quad 0 \leftarrow t / \tau_{UV,\Lambda} \rightarrow \infty ,
\end{array}
\ee
where ${1 \over R_I ^2} \equiv {8 \pi G_N \rho_I \over 3 c^4}$ and
${1 \over R_\Lambda ^2} \equiv {8 \pi G_N \rho_\Lambda \over 3 c^4}$.
During the early thermalization of
any initial state quantum condensate, the cosmological
quantum scale $R_v$ is expected
to rapidly vary from the initial vacuum scale $R_I$ to the final
deSitter scale $R_\Lambda$.  The general dynamics of
the cosmology will be explored in reversed temporal order.

\setcounter{equation}{0}
\section{Present to Late Time Cosmology}
\indent

For the period from matter domination to dark energy domination,
the parameters are chosen to satisfy
$\rho_v \cong \rho_\Lambda$, $R_v \cong R_\Lambda$, and $w=0$.
Eq. \ref{thermalrhodynamics} takes the form
\be
\begin{array}{l}
{d \over d ct} \rho_{thermal} \cong -\sqrt{{24 \pi G_N   \over c^4 } (\rho_\Lambda +
\rho_{thermal}) } \, \rho_{thermal}  \\ \\
{  \sqrt{\rho_{thermal} + \rho_\Lambda} - \sqrt{\rho_\Lambda}   \over  
 \sqrt{\rho_{thermal} + \rho_\Lambda} + \sqrt{\rho_\Lambda}    } =
{  \sqrt{\rho_* + \rho_\Lambda} - \sqrt{\rho_\Lambda}   \over  
 \sqrt{\rho_* + \rho_\Lambda} + \sqrt{\rho_\Lambda}    } \quad
e^{-{3 \over R_\Lambda} (ct - ct_*)} .
\end{array}
\label{vacuumdensity}
\ee
For brevity, define the ratio at time $t_*$ as
$A_* \equiv {  \sqrt{\rho_* + \rho_\Lambda} - \sqrt{\rho_\Lambda}   \over  
 \sqrt{\rho_* + \rho_\Lambda} + \sqrt{\rho_\Lambda}    }$.  Using
Eq. \ref{FRWdSdensity}, the reduced scale factor can be calculated
as
\be
{\mathcal{R}(ct) \over \mathcal{R}_*} = \left [
{  e^{3ct/2 R_\Lambda} - A_* e^{-3ct/2 R_\Lambda} \over
e^{3ct_*/2 R_\Lambda} - A_* e^{-3ct_*/2 R_\Lambda} }
\right ]^{2/3} ,
\ee
whereas the thermal scale factor satisfies
\be
{R(ct) \over R_\infty } =  \left [ 
1 - A_* e^{- 3 ct /R_\Lambda} \right ] ^{2/3} .
\ee
For consistency, the thermal scale corresponds to the horizon
scale at late times $R_\infty =R_\Lambda$, which gives the
thermal scale at $t_*$ as
${R_* \over R_\Lambda}=(1-A_* e^{-3 ct_*/R_\Lambda})^{2/3}$.  Writing the Hubble
rate for the reduced scale parameter
at a time $t_*$ during the matter dominated
epoch as ${ \dot{\mathcal{R}}_*  \over \mathcal{R}_*}
\cong {\dot{R}_* \over R_*}
\equiv {H_* \over c}$, the ratio $A_*$ can be re-expressed
using Eq. \ref{FRWdSdensity}
as $A_* \cong 1/(1+ 2{c \over H_* R_\Lambda })$.  Therefore, the
thermal scale at matter-radiation equality can be 
estimated as
\be
R_{eq} \cong \left ( 3 ct_{eq} + {2 c \over H_{eq}} \right ) ^{2/3} R_\Lambda ^{1/3}.
\label{lateequality}
\ee

\setcounter{equation}{0}
\section{Intermediate (Classical) Evolution}
\indent

During the intermediate (thermal) period $\tau_{UV} << t << \tau_\Lambda $, one
assumes that $\rho_v << \rho_{thermal}$ and
$\left | {\dot{R} \over R } \right |  >> {1 \over R_v} \cong {1 \over R_\Lambda}
\rightarrow {\dot{\mathcal{R}} \over \mathcal{R}} \cong {\dot{R} \over R}$.  The
thermal density satisfies
\be
\begin{array}{l}
{d \over d ct} \rho_{thermal} = -\sqrt{{24 \pi G_N   \over c^4 } } 
(1+w_*) \rho_{thermal} ^{3/2} \\ \\
\left(  {\rho_* \over \rho_{thermal}} \right ) ^{1/2} \cong
1 + \sqrt{{6 \pi G_N \rho_*  \over c^4 } } (1+w_*) (ct - ct_*) ,
\end{array}
\label{thermaldensity}
\ee
which are the same as the behaviors predicted by the
Friedman-Lemaitre equations during the
thermal period\cite{BigBang}.
Using Eq. \ref{FRWdSdensity}, the forms of the reduced
scale parameter can be determined:
\be
{\mathcal{R} \over \mathcal{R}_*} =
\left [
1+{3 \over 2} (1 + w_*) {\dot{\mathcal{R}}_* \over \mathcal{R}_* }
(ct - ct_*) \right ] ^ {{2 \over 3 (1+w_*)}} ,
\ee
where again  ${ \dot{\mathcal{R}}_*  \over \mathcal{R}_*}
\cong {\dot{R}_* \over R_*} = {H_* \over c}$.  
During these epochs, the reduced and thermal scale parameters
are related by
\be
{R \over R_*} =
{\mathcal{R} \over \mathcal{R}_* }
 e^{- {ct-ct_* \over R_\Lambda}} .
\ee
The thermal scale ratio is seen to differ significantly from
the reduced scale ratio only when the time is comparable to
that required for light to traverse the deSitter horizon scale 
for the dark energy $R_\Lambda$.

\setcounter{equation}{0}
\section{Early Time Evolution}
\indent

An initial quasi-stationary quantum state is likely in an
energy ground state.
One can argue that the temporal dynamics begins when the
energy content of the cosmology has a non-vanishing
thermal fraction\cite{NSBP06}.  Therefore, the initial state $t=0$ will
be taken to have a vanishing thermal fraction.
Since the cosmology thermalizes into a radiation dominated
epoch, the dynamics of density evolution Eq. \ref{thermalrhodynamics}
with $w={1 \over 3}$ gives
\be
{d \over d ct} \rho = -  {4 \over 3} \rho_{thermal}
\sqrt{{24 \pi G_N  \over c^4 } \rho } .
\label{earlyrhodynamics}
\ee
It is convenient to define the thermal fraction
\be
f(ct) \equiv {\rho_{thermal} \over \rho}.
\ee
The dynamics of the thermal fraction
$f(ct)$ gives microscopic physical input for the
early cosmology, and sets the early scales
of the system.  Its temporal behavior is expected to
satisfy $0 \Leftarrow f(ct)   \Rightarrow 1-{\rho_\Lambda \over
\rho_{UV}}$ for  $0 \leftarrow ct \rightarrow ct_{UV}$, where
the density $\rho_{UV}$ represents the thermal energy
density of the radiation dominated universe immediately
following thermalization.  This
allows the density in Eq. \ref{earlyrhodynamics} to be
determined in the form
\be
\left ( {\rho_I \over \rho} \right ) ^{1/2} =
1 + 2 \sqrt{{8 \pi G_N \rho_I \over 3 c^4}}
\int _0 ^{ct} f(ct') dct' =
1 + {2 \over R_I}
\int _0 ^{ct} f(ct') dct' ,
\label{earlydensity}
\ee
which smoothly connects to the behavior
expected during the thermal epoch Eq. \ref{thermaldensity}
for $t>t_{UV}$ in the form
\be
\left ( {\rho_I \over \rho} \right ) ^{1/2} =
{\mathcal{R}\over \dot{\mathcal{R}} R_I} \cong
1 + {2 \over R_I} \left [
\int _0 ^{ct_{UV}} f(ct') dct' +(ct - ct_{UV}) \right ] .
\ee
This equation relates the density at
radiation-matter density equality to the time $t_{eq}
>> t_{UV}$:
\be 
{8 \pi G_N \over 3 c^4} \rho_{eq} \cong  \left ( 
{1 \over  2 ct_{eq}} \right ) ^2 . 
\ee
The general solution for the reduced scale $\mathcal{R}$
using Eq. \ref{earlydensity} is given by 
\be
{\mathcal{R} \over R_I} = exp \left [
\int _0 ^{ct}{cdt'' \over R_I + 2 \int_0 ^{ct''} f(ct') \, dct'}
\right ] .
\label{earlythermalscale}
\ee
Assuming $f(ct>ct_{UV}) \cong 1$, the
solution takes the form
\be
{\mathcal{R} \over \mathcal{R}_{UV}} \cong \left [
1 + {2 (ct - ct_{UV}) \over R_I + 2 \int_0 ^{ct_{UV}} f(ct') \, dct'}
\right ] ^{1/2} ,
\ee
which is the expected form for the time evolution of the
cosmological scale during radiation domination.

Now that correspondence with the observed behavior of the
early thermal universe has been established, the useful
early time behavior of the model will be explored.  The condition
of a unique initial cosmological scale $R_I$ with vanishing initial
expansion rate for the thermal scale $\dot{R}(0)=0$
implies that $\dot{\mathcal{R}}(0)=1$ corresponds to the onset
of thermalization.  The ``vacuum" scale $R_v$ is determined by
the ``vacuum energy density" using
\be
{1 \over R_v ^2} \equiv 
{8 \pi G_N \over 3 c^4} \rho_v = {8 \pi G_N \over 3 c^4} (1-f) \rho .
\ee
From Eq. \ref{FRWdSdensity}, the temporal behaviors of
the scales during thermalization can be related:
\be
\begin{array}{l}
{1 \over R_v} = {(1-f) + \sqrt{1-f} \over f} {\dot{R} \over R}, \\ \\
{\dot{\mathcal{R}} \over \mathcal{R}} =
{ 1 + \sqrt{1-f} \over f} {\dot{R} \over R} .
\end{array}
\label{fractionalrates}
\ee
During thermalization, the thermal expansion rate ${\dot{R} \over R}$
increases from 0 to its maximum value $\sqrt{{ 8 \pi G_N \rho_{UV}
\over 3 c^4}}$, and the
the reduced expansion rate ${\dot{\mathcal{R}} \over \mathcal{R}}$
is seen to decrease from its maximum value ${1 \over R_I}$
at time $t=0$ to essentially the same as the 
thermal scale expansion rate.
The relative contributions of the scales $R_v$ and $R$ to the
expansion rate of the reduced scale $\mathcal{R}$ become equal
when the thermal fraction takes the value $f={3 \over 4}$.

Using Eq. \ref{fractionalrates} and \ref{earlythermalscale},
the dynamics of the thermal scale during thermalization
can be determined:
\be
log \left ( {R(ct) \over R_I} \right ) =
\int _0 ^ {ct}  {c dt'' \over R_I + 2 \int _0 ^{ct''} f(ct') \, dct'}
\left ( {f(ct'') \over 1 + \sqrt{1 - f(ct'')}} \right ).
\ee
This form directly relates the later time cosmological
parameters Eq. \ref{lateequality} with the early time behavior
\be
\left ( {R(ct_{eq}) \over R_I} \right ) \cong 
{1 \over R_I} \left ( 3 ct_{eq} +{2 c \over H_{eq}} \right ) ^{2/3} R_\Lambda ^{1/3} .
\ee
Further exploration requires a microscopic model that
describes the energy density fractions of thermal and vacuum
energies during the period of significant macroscopically coherent content.

\setcounter{equation}{0}
\section{Conclusions}
\indent

A consistent model of a universe described by a perfect fluid with
no cosmological constant has been developed.  The model
allows for the dynamical evolution of a vacuum energy scale
along with a more classically behaved thermal energy scale.

During early times, the classical dynamics is generated
by a non-vanishing fraction of thermal energy content, eventually
dominating the cosmological expansion during the intermediate
period of the expansion of the universe.  Particular microscopic physical
models specify the detailed temporal dependence of this
thermal fraction.

Correspondence of the behavior of the thermal scale
with that in the Friedman-Lemaitre equations in standard
cosmology has been demonstrated during the
intermediate evolution of the model.  In addition,
the behavior in late times is consistent with
dark energy domination as would occur in a
cosmology with an additional cosmological constant
term added to Einstein's equation.

A transformation that compares the coordinates
and scales of the metric Eq. \ref{FRWdSmetric} with those of a
Friedman-Robertson-Walker space-time
will be presented in future work.

\end{document}